\documentclass{article}
\usepackage{spconf,amsmath,graphicx,hyperref}

\usepackage{array}
\usepackage{booktabs}
\usepackage{xcolor}
\usepackage{svg} 

\title{OneVoice: An Intermediate Representation for Agentic Speech Pipelines}
\name{Vipul Charugundla, Dancheng Liu, Jinjun Xiong \thanks{Thanks to XYZ agency for funding.}}
\address{University at Buffalo,  University of Texas at San Antonio}
\begin{document}
%
\maketitle
\begin{abstract}
Agentic speech systems must exchange more than text, including speaker identity, timing, phonetic information, and behavioral annotations. Yet these signals are often produced in incompatible tool-specific formats, making agent-to-agent handoff fragile. We present OneVoice, a lightweight, JSON-native intermediate representation that provides a shared semantic structure for speech pipelines. OneVoice organizes heterogeneous speech evidence into validated session records with stable identifiers, layered transcripts, explicit timing relationships, and provenance information. We evaluate OneVoice in two complementary multi-agent workflows covering event aggregation and acoustic-phonetic temporal linking across three language models. Compared with implicit agent-defined handoffs, OneVoice substantially improves aggregation reliability and the preservation of temporal relationships, demonstrating the value of an explicit speech-specific representation for agent communication.
\end{abstract}

\section{Introduction}

Agentic AI systems increasingly operate as multi-step pipelines in which agents invoke specialized tools, interpret their outputs, and pass intermediate results downstream~\cite{allbert2025evaluating,lee2025performance}. Speech makes this interaction particularly challenging. A spoken utterance carries much more than words alone. Useful evidence may include who spoke, when each word or phone occurred, how something was pronounced, and what behavioral or interactional events were present. These signals are often produced separately by ASR, diarization, forced alignment, and other speech analysis tools~\cite{povey2011kaldi,bredin2020pyannote,bain2023whisperx,eyben2010opensmile}. As a result, a successful speech-agent pipeline depends not only on the accuracy of its individual components, but also on whether the information they produce can survive the handoff from one agent to the next.

\begin{figure}[t!]
\centering
\includegraphics[width=0.48\textwidth]{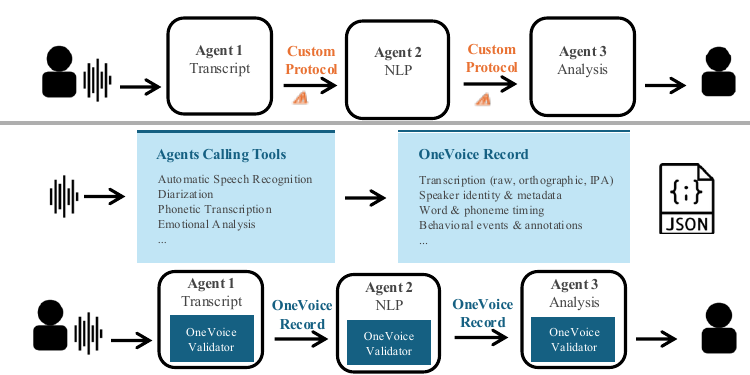}
\caption{Speech-agent communication with pairwise, tool-specific handoffs in the conventional workflow and a shared, validated OneVoice representation in the proposed workflow.}
\label{fig:onevoice_comparison}
\vspace{-4mm}
\end{figure}

Such handoffs are difficult because today's speech artifacts were largely designed for individual tools, datasets, or annotation environments rather than for communication between agents. Timing information may appear in TextGrid or alignment files~\cite{boersma2001praat,mcauliffe2017montreal}, interactional annotations in CHAT transcripts~\cite{macwhinney2000childes}, and other outputs in text, CSV, or ad hoc JSON. Each representation works well within its original setting, yet combining them often requires custom parsing and implicit assumptions about how transcripts, speakers, timestamps, and annotations correspond~\cite{fuentes2021soundata}. These assumptions become fragile as information moves through a pipeline. Transcript variants may overwrite one another, timing may become detached from the speech units it describes, and missing annotations may be confused with events that were checked and found absent. Figure~\ref{fig:onevoice_comparison} illustrates this problem. Without a shared representation, every handoff becomes another opportunity to reinterpret or lose speech-specific structure.

We introduce \textbf{OneVoice}, a lightweight, JSON-native intermediate representation designed to make these handoffs explicit and consistent. Rather than replacing existing speech formats, OneVoice acts as a runtime interchange layer between them. It organizes heterogeneous speech evidence into a shared session record while preserving three forms of structure that are easily lost across tools. Stable identifiers provide join keys across independently generated artifacts. Layered representations keep transcript variants, speaker information, and optional word- and phone-level timing connected without forcing one view to replace another. Explicit provenance and annotation coverage distinguish unavailable information from partial or observed annotations. A validator further checks structural, temporal, and referential consistency before records propagate through the pipeline.

We evaluate OneVoice in two complementary multi-agent speech workflows that stress different forms of information preservation across agent handoffs. 
Across three large language models, OneVoice substantially improves handoff fidelity in both settings, reducing unstable aggregation and improving temporally grounded linking. These results show that reliable speech-agent handoff depends on preserving the structure and relationships within speech evidence, not simply exchanging it in a common format.
\section{Related Work}

Speech information is represented across several tool- and task-specific ecosystems. Praat TextGrid and word/phone alignment files encode temporal structure~\cite{boersma2001praat,garofolo1993timit}, while formats such as CHILDES/CHAT capture rich interactional annotations~\cite{macwhinney2000childes}. Forced aligners, diarization systems, and other speech tools introduce further task-specific outputs~\cite{mcauliffe2017montreal}. These representations are effective within their intended workflows, but their differing structures and identifiers make information exchange across tools difficult.

More general annotation frameworks have long addressed the need to represent layered linguistic information. Annotation Graphs model annotations over time-series data~\cite{bird2001formal}, while LAF/GrAF, ELAN/EAF, NITE, and EXMARaLDA provide structured representations for multi-layer, temporally aligned, and cross-referenced annotations~\cite{ide2004international,ide2007graf,wittenburg2006elan,carletta2003nite,schmidt2014transcribing}. These systems provide important foundations for representing complex speech annotations. Their primary focus, however, is corpus annotation, expert editing, or general-purpose interchange, often through XML-based representations. OneVoice builds on these ideas but targets a different setting in which LLM agents must repeatedly exchange, validate, and combine speech evidence during pipeline execution.

Other efforts address complementary layers of interoperability. Soundata and HuggingFace Datasets standardize access to heterogeneous datasets while largely preserving dataset-specific annotation structures~\cite{fuentes2021soundata,lhoest2021datasets}. Agent protocols such as MCP and A2A standardize tool invocation and message exchange, but do not define how speech-specific content such as transcripts, speakers, timing, and annotations should be represented~\cite{hou2025model,Surapaneni2025A2A}. OneVoice targets this missing content layer by providing a lightweight, validated speech representation with stable cross-artifact identifiers and explicit relationships among speech evidence.
\section{Proposed Schema}

\begin{figure}[t]
\centering
\includegraphics[width=0.47\textwidth]{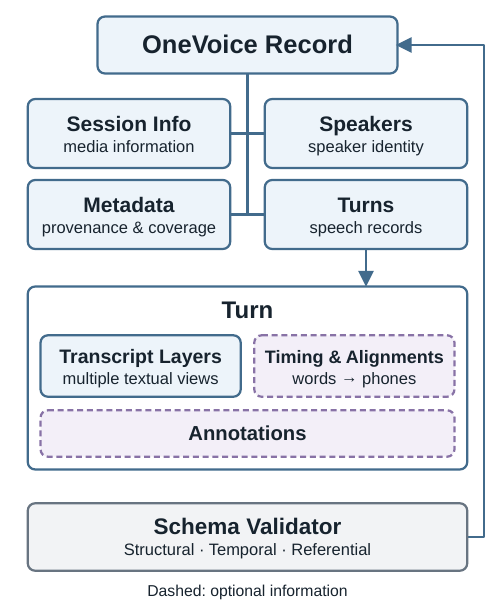}
\caption{High-level structure of a OneVoice record. A session contains speaker and metadata blocks together with turn-level records. Each turn may include multiple transcript layers, optional word- and phone-level alignments, and optional task-specific annotations.}
\label{fig:onevoice_schema}
\vspace{-4mm}
\end{figure}

OneVoice is a lightweight, JSON-native intermediate representation for runtime exchange between speech agents and tools. It is not intended to replace existing annotation or dataset formats. Instead, it showcases a common handoff structure that lets agents exchange speech evidence without repeatedly reconstructing how transcripts, speakers, timing, and annotations correspond. Figure~\ref{fig:onevoice_schema} shows the high-level organization of a OneVoice record. Detailed field definitions, validation rules, and examples are provided in the GitHub repository \footnote{\url{https://github.com/Charugundlavipul/OneVoice}}.

The design of OneVoice centers on three ideas. First, OneVoice uses stable identifiers for sessions, speakers, and turns, together with explicit hierarchical references for words and phones, so that independently produced artifacts can be joined without relying on filenames or implicit ordering conventions. Second, it preserves multiple views of the same utterance through explicit transcript layers, allowing raw source text, cleaned orthography, reference prompts, and phonetic or IPA representations to coexist without overwriting one another. Third, it records provenance and annotation coverage explicitly, allowing downstream agents to distinguish unavailable information from partial or observed annotations.

A OneVoice record is organized around a session and its associated turns. The session stores media-level information together with speaker and metadata blocks, while each turn serves as the basic integration unit for transcript layers, optional timing and word- and phone-level alignments, and optional task-specific annotations such as mispronunciations or behavioral events. To accommodate heterogeneous speech resources, fields that permit empty values may remain unpopulated when information is unavailable. For text and phoneme annotations, provenance and coverage metadata clarify the availability and completeness of the recorded evidence, so that empty values are not mistaken for observed absences.

During pipeline execution, agents operate on the same logical session record rather than exchanging isolated tool outputs. An upstream agent may populate or extend the portions of the record relevant to its task, while downstream agents can retrieve those fields through stable references without reconstructing their correspondence from filenames, ordering, or free-form text. Existing evidence is preserved as new annotations are added, allowing multiple tools to contribute complementary views of the same speech segment. 

To support reliable agent handoff, OneVoice is paired with a schema validator. Before the updated record is forwarded, the validator checks that newly added information remains consistent with the existing session structure. Beyond checking syntactic validity, the validator also enforces relationships that would otherwise remain implicit across independently produced artifacts. It verifies valid temporal ranges, containment of word and phone intervals within their parents, and consistency of references such as speaker identifiers. When validation fails, the offending record can be returned to the producing agent for repair before the error propagates downstream. Dataset- or tool-specific information remains extensible through namespaced fields in the optional annotations without changing the core representation.
\section{Experiments and Results}

\subsection{Experimental Setup}

We evaluate whether OneVoice improves the reliability of structured speech information as it passes between agents. Both workflows follow the same three-agent design to simulate a minimal multi-agent handoff scenario. Agents A and B independently analyze different aspects of the input, while Agent C combines their outputs into a fixed final evaluation format. This keeps the downstream task unchanged while varying only the representation used during agent handoff.

We compare three communication conditions. In \textbf{C0} (\emph{implicit}), agents define their own intermediate JSON representations and are instructed only to provide information useful to the next agent. In \textbf{C1} (\emph{OneVoice}), intermediate outputs follow the OneVoice schema but are passed without repair. In \textbf{C2} (\emph{OneVoice+validator}), records are validated after Agents A and B, and detected errors are returned to the model for up to three repair rounds before the record reaches Agent C. All experiments use temperature zero and are evaluated with \texttt{GPT-5 mini}, \texttt{GPT-5.4 mini}, and \texttt{Claude Haiku 4.5}.

The OneVoice toolkit additionally provides converters for common speech resources, including CHAT, TextGrid, RTTM, plain-text, tabular, and HuggingFace-style inputs. The same GitHub repository provides full prompts, conversion procedures, and implementation details.

\subsection{Tasks}

\textbf{UXSSD event aggregation.}
We use 15 child-speech bundles from the Ultrasuite
dataset~\cite{eshky18_interspeech}, each containing approximately
2--3 minutes of contiguous speech. Gold annotations contain
94 mispronunciation events and 112 behavioral events. Agent A
extracts mispronunciations, Agent B extracts behavioral events,
and Agent C combines their outputs into final event counts.
This task evaluates count preservation across agent handoffs.

\textbf{TIMIT temporal linking.}
We use 40 TIMIT speaker-level bundles, each containing five
utterances from the same speaker~\cite{garofolo1993timit}.
Gold events and temporal relationships are derived
deterministically from \texttt{.WRD} and \texttt{.PHN} annotations.
Agent A identifies word-duration outliers and inter-word gaps;
Agent B identifies phone-duration outliers, closures, silence,
and glottal stops. Agent C must preserve both event sets and
link phone-level events to parent words using temporal
containment or overlap. This task evaluates event aggregation
together with temporal and hierarchical consistency.

\subsection{Metrics}

For both tasks, we report precision, recall, $F_1$, and mean absolute error (MAE) over event counts. Because the evaluation concerns event inventories rather than boundary-matched detections, we use a count-based definition. For each evaluation unit $i$ and event type $e$, the number of matched events is
\begin{equation}
    \mathrm{TP}_{i,e}
    = \min(\mathrm{gold}_{i,e}, \mathrm{pred}_{i,e}),
\end{equation}
after which precision, recall, and $F_1$ are micro-averaged over all units and event types.

For TIMIT, word-link accuracy measures the fraction of evaluated
phone-level events whose predicted parent word agrees with the
gold annotation. Invalid-link rate measures the fraction of all
predicted phone-level events whose parent-word references do not
resolve to valid words. These metrics have different denominators,
and a valid reference may still identify an incorrect word.
Boundary error measures the mean absolute deviation of predicted
start and end times from the corresponding gold boundaries,
averaged over evaluated events and reported in milliseconds.
\subsection{Results}

\begin{table*}[t]
\centering
\setlength{\tabcolsep}{17pt}
\renewcommand{\arraystretch}{1.05}
\caption{UXSSD event aggregation results. }
\label{tab:uxssd-cross-model}
\begin{tabular}{llcccc}
\toprule
Model & Condition &
Precision $\uparrow$ &
Recall $\uparrow$ &
$F_1$ $\uparrow$ &
MAE $\downarrow$ \\
\midrule
\texttt{GPT-5 mini}
& C0 implicit
& 0.156 & \textbf{0.874} & 0.265 & 66.47 \\
& C1 OneVoice
& \textbf{0.933} & 0.811 & 0.868 & \textbf{3.40} \\
& C2 OneVoice+val.
& 0.877 & 0.869 & \textbf{0.873} & 3.47 \\
\midrule
\texttt{GPT-5.4 mini}
& C0 implicit
& 0.264 & \textbf{0.888} & 0.407 & 35.60 \\
& C1 OneVoice
& 0.607 & 0.840 & 0.705 & 9.67 \\
& C2 OneVoice+val.
& \textbf{0.886} & 0.714 & \textbf{0.790} & \textbf{5.20} \\
\midrule
\texttt{Claude Haiku 4.5}
& C0 implicit
& 0.574 & 0.340 & 0.427 & 12.53 \\
& C1 OneVoice
& 0.656 & 0.937 & 0.772 & 7.60 \\
& C2 OneVoice+val.
& \textbf{0.687} & \textbf{0.947} & \textbf{0.796} & \textbf{6.67} \\
\bottomrule
\end{tabular}
\vspace{-3mm}
\end{table*}

\begin{table*}[th]
\centering
\setlength{\tabcolsep}{2.5pt}
\caption{TIMIT temporal linking results. Word-link accuracy evaluates hierarchical association between phone- and word-level evidence, while boundary error measures temporal localization.}
\label{tab:timit-results}
\begin{tabular}{llcccccc}
\toprule
Model & Condition &
Precision $\uparrow$ &
Recall $\uparrow$ &
$F_1$ $\uparrow$ &
Word-link acc. $\uparrow$ &
Invalid link $\downarrow$ &
Boundary err. $\downarrow$ \\
\midrule
\texttt{GPT-5 mini}
& C0 implicit
& 0.908 & 0.800 & 0.851 & 
0.819 & \textbf{0.164} & 194.45 ms \\
& C1 OneVoice
& \textbf{0.978} & 0.823 & 0.894 & 
0.864 & 0.205 & 155.56 ms \\
& C2 OneVoice+val.
& 0.969 & \textbf{0.906} & \textbf{0.936} & 
\textbf{0.945} & 0.201 & \textbf{99.63 ms} \\
\midrule
\texttt{GPT-5.4 mini}
& C0 implicit
& 0.542 & \textbf{0.604} & 0.571 & 
0.595 & \textbf{0.192} & 478.61 ms \\
& C1 OneVoice
& \textbf{0.762} & 0.345 & 0.475 & 
\textbf{0.765} & 0.367 & 415.14 ms \\
& C2 OneVoice+val.
& 0.618 & 0.558 & \textbf{0.586} & 
0.755 & 0.204 & \textbf{352.35 ms} \\
\midrule
\texttt{Claude Haiku 4.5}
& C0 implicit
& 0.668 & 0.589 & 0.626 & 
0.586 & 0.204 & 571.98 ms \\
& C1 OneVoice
& \textbf{0.805} & 0.431 & 0.561 & 
0.629 & 0.213 & 549.04 ms \\
& C2 OneVoice+val.
& 0.759 & \textbf{0.638} & \textbf{0.693} & 
\textbf{0.730} & \textbf{0.160} & \textbf{431.29 ms} \\
\bottomrule
\end{tabular}
\vspace{-3mm}
\end{table*}

On UXSSD, OneVoice improves the stability of event aggregation across all three models. The most obvious example is \texttt{GPT-5 mini}, for which implicit handoff produces 1,151 predicted events against only 206 gold events, resulting in an $F_1$ of 0.265 and an MAE of 66.47. With OneVoice, the over-generation is largely eliminated, raising $F_1$ to 0.868 in C1 and 0.873 in C2 while reducing MAE to 3.4. The same pattern appears for \texttt{GPT-5.4 mini} and \texttt{Claude Haiku}, whose best $F_1$ scores increase from 0.407 to 0.790 and from 0.427 to 0.796, respectively. These results show that an explicit handoff structure can prevent independently generated evidence from becoming unstable during aggregation.

TIMIT stresses a different form of handoff fidelity. Here, agents must preserve not only detected events but also their temporal and hierarchical relationships. For \texttt{GPT-5 mini}, C2 raises word-link accuracy from 0.819 to 0.945 and reduces boundary error from 194.45 to 99.63~ms, while event $F_1$ improves from 0.851 to 0.936. \texttt{Claude Haiku} shows the same overall pattern, with word-link accuracy increasing from 0.586 to 0.730 and boundary error decreasing from 571.98 to 431.29~ms. For \texttt{GPT-5.4 mini}, C2 improves $F_1$ and reduces boundary error over implicit handoff, while C1 gives the highest word-link accuracy. Across models, the results indicate that explicit temporal and referential structure helps preserve relationships that are otherwise left implicit during agent communication.

Taken together, the two tasks expose complementary failure modes. UXSSD shows how loosely structured handoffs can lead to severe aggregation errors, while TIMIT shows how temporal and parent--child relationships can degrade as evidence moves between agents. OneVoice improves both forms of handoff fidelity, and validator-in-the-loop repair provides additional gains when structural inconsistencies can be detected before they propagate downstream.
\section{Conclusion}

We presented OneVoice, a lightweight intermediate representation for agentic speech pipelines that organizes heterogeneous speech artifacts into validated session records with stable identifiers, layered transcripts, explicit timing relationships, and provenance information. Across two multi-agent workflows, OneVoice improves handoff fidelity by reducing unstable event aggregation and better preserving temporal and hierarchical relationships. These results demonstrate the value of a shared speech-specific representation for reliable communication across agentic speech pipelines.

\bibliographystyle{IEEEbib}
\bibliography{strings,refs}

\end{document}